\documentclass[12pt,onecolumn,canadian,english,journal]{IEEEtran}
\usepackage[T1]{fontenc}
\usepackage[latin9]{inputenc}
\usepackage{amsmath}
\usepackage{amsthm}
\usepackage{amssymb}
\usepackage{graphicx}

\makeatletter
\theoremstyle{plain}
\newtheorem{thm}{\protect\theoremname}
\theoremstyle{definition}
\newtheorem{defn}[thm]{\protect\definitionname}
\theoremstyle{plain}
\newtheorem{lem}[thm]{\protect\lemmaname}
\theoremstyle{plain}
\newtheorem{cor}[thm]{\protect\corollaryname}
\theoremstyle{remark}
\newtheorem{rem}[thm]{\protect\remarkname}

\makeatother

\usepackage{babel}
\addto\captionscanadian{\renewcommand{\corollaryname}{Corollary}}
\addto\captionscanadian{\renewcommand{\definitionname}{Definition}}
\addto\captionscanadian{\renewcommand{\lemmaname}{Lemma}}
\addto\captionscanadian{\renewcommand{\remarkname}{Remark}}
\addto\captionscanadian{\renewcommand{\theoremname}{Theorem}}
\addto\captionsenglish{\renewcommand{\corollaryname}{Corollary}}
\addto\captionsenglish{\renewcommand{\definitionname}{Definition}}
\addto\captionsenglish{\renewcommand{\lemmaname}{Lemma}}
\addto\captionsenglish{\renewcommand{\remarkname}{Remark}}
\addto\captionsenglish{\renewcommand{\theoremname}{Theorem}}
\providecommand{\corollaryname}{Corollary}
\providecommand{\definitionname}{Definition}
\providecommand{\lemmaname}{Lemma}
\providecommand{\remarkname}{Remark}
\providecommand{\theoremname}{Theorem}

\begin{document}
\title{Merged Bitcoin: Proof of Work Blockchains with Multiple Hash
Types}
\author{Christopher Blake, Chen Feng, Xuachao Wang, Qianyu Yu}
\maketitle
\begin{abstract}
Proof of work blockchain protocols using multiple hash types are considered.
It is proven that the security region of such a protocol cannot be
the AND of a 51\% attack on all the hash types. Nevertheless, a protocol
called Merged Bitcoin is introduced, which is the Bitcoin protocol
where links between blocks can be formed using multiple different
hash types. Closed form bounds on its security region in the $\Delta$-bounded
delay network model are proven, and these bounds are compared to simulation
results. This protocol is proven to maximize cost of attack in the
linear cost-per-hash model. A difficulty adjustment method is introduced,
and it is argued that this can partly remedy asymmetric advantages
an adversary may gain in hashing power for some hash types, including
from algorithmic advances, quantum attacks like Grover's algorithm,
or hardware backdoor attacks.
\end{abstract}

Since it was introduced in 2008 \cite{Nakamoto_bitcoin}, the Bitcoin
protocol has become a global database securing over one trillion USD
in value \cite{CoinMarketCap}. In \cite{bitcoinAndGold},
the author proposes that an adversary could use financial instruments
to make sufficiently high profit to cover the cost of a short term
$51\%$ attack. Moreover, some have proposed that Bitcoin can become
a world reserve currency that replaces or complements gold in financial
markets, which would further increase the potential reward of a $51\%$
attack \cite{bitcoinStandard}. Thus the primary motivation of this paper is to increase the security of the Bitcoin protocol, to alleviate the risk of such catastrophic attacks. To increase security, the standard
way is to increase honest mining power.

Currently, economical mining of Bitcoin requires using specialized
circuits that compute the SHA-256 hashing function. Thus, an obvious
way to increase the security threshold is to increase the number of
these circuits. However, such an approach has vulnerabilities. First,
increasing such mining power further centralizes the protocol to agents
with the best access to these circuits. Second, this approach has
the possibility of a hardware backdoor attack, where the limited number
of specialists producing Bitcoin mining ASICs could be compromised,
leaving the Bitcoin protocol vulnerable even if the majority of miners
are honest. 

In this paper, we propose a new approach to increase honest mining
power called \emph{Merged Bitcoin.} Simply put, this protocol allows
blocks linked by different hash types to form a chain. Otherwise,
it is exactly the Bitcoin protocol. This increases the security threshold,
and it reduces the risk of the protocol being centralized to a few
agents that specialize in mining with SHA-256 circuits, and could
allow personal computer to viably participate in the protocol.\footnote{The bottleneck for SHA-256 is specialized circuits. But the bottleneck
of different hashes are different computational resources. For example,
before it was converted to Proof of Stake, Ethereum used Ethhash,
whose bottleneck was memory-access bandwidth \cite{ethereumDesignJustificationWiki}.
Such hashes can be economically computed by personal computers, and
do not require specialized circuitry to viably compete.} As well, it reduces the risk of a hardware backdoor attack or quantum
attack specialized to one specific hash type.

We consider the $\Delta$-bounded delay network model. In a companion
paper \cite{correctionPaper}, we prove that the mining power required to break Merged
Bitcoin is the power required to mine faster than the honest, fully-delayed score growth rate. This reduces to the
sum of the mining powers of each hash type when delay approaches $0$.
In contrast to the companion paper, the technical focus of this paper
is on bounding the fully-delayed score growth rate. This gives us
closed form upper and lower bounds on the security region of the protocol.

The motivation of this paper comes from an information theoretic idea:
that if security can be broken only when an AND of different 51\%
thresholds are met, then an information-theoretic exponentially-decreasing
error probability can be achieved. Unfortunately, we show that this
is \emph{not }possible for \emph{any} permissionless distributed protocol
(and not just Merged Bitcoin). We prove this using a simulation argument
which is similar to other proofs in the literature \cite{minotaurFitziEtAl}.
The key issue is permissionlessness: once this is allowed, any adversary
can simply simulate honest mining. Nevertheless, our proposed protocol
does increase the \emph{cost }of an attack, which we describe formally.
We introduce a linear cost-per-hash economic model and prove that
with appropriate choice of constants Merged Bitcoin is optimal in
the case of negligible network delay. As well, as we will discuss,
using $n$ different hash types can decrease the probability of a
hardware backdoor attack exponentially in $n$, with some natural
independence assumptions on probability of attacking one block type.
Thus, our information theoretic motivation bears some fruit.

\section{Prior Work}\label{sec:Prior-Work}

 Prior work in \cite{judmayer2017merged} introduces the concept of \emph{merged mining}. This is not to be confused with our algorithm, \emph{Merged Bitcoin}.  Merged \emph{mining} involves
using one blockchain (say, Bitcoin), to act as a consensus layer for
a \emph{different} distributed ledger, hence gaining more utility
out of the parent blockchain, and securing the other ledger. Our work
is distinguished from this; our goal is \emph{not }to use one blockchain
to secure a different ledger; rather it is to merge different hashing
resources to produce a single, more secure ledger.

There has been some work on protocols that involve merging resources.
Early work on merging proof-of-work and proof-of-stake occurred in
\cite{bentovProofOfActivity}. In Fitzi\emph{ et al. }\cite{minotaurFitziEtAl},
the authors consider a way to merge work \emph{and }stake for a protocol
called Minotaur. Fundamentally, Minotaur involves a proof-of-stake
backbone, where proof-of-work mining is done to establish rights to
the proof-of-stake mining protocol. The authors of \cite{twinsCoinChepurnoy}
also propose a protocol merging stake and work.

In contrast to the Minotaur protocol, our protocol has different security
assumptions. Since Minotaur is a merged protocol using a proof-of-stake
backbone, and it claims similar security regions as the Merged Bitcoin
protocol we propose in this paper, we discuss the tradeoffs between
the two approaches in Subsection \ref{subsec:Comparison-of-Minotaur}.

\section{The Multiple Random Resource Model}

We assume that nodes, honest or dishonest, have \emph{hashing resources},
which could vary in amount. These resource amounts parameterize a
Poisson arrival process. Underlying this is a cryptographic assumption
that these hashes can be used to create an ideal source of randomness.

Specifically, hashing contests can be run by having all nodes with
hashing resource $i$ compute \\
$hash_{i}(msg,nonce)$. The message
($msg$), can include transactions, timestamps, hashes of prior blocks,
or anything else as required by the protocol. The nonce is some random
number that the node chooses as a ``guess.'' If the value of the
hash is less than some threshold value, it is considered \emph{valid.}
This process produces a sequence of valid hashes whose arrival times
form a Poisson process.

Let $q_{i}^{h}$ and $q_{i}^{a}$ be the honest and adversary hashrates
for resource $i$. These are measured in hashes per second. Let $d_{i}$
be the difficulty levels for hash type $i$. These can be measured
in units of hashes per block.

We define
\[
h_{i}=\frac{q_{i}^{h}}{d_{i}}
\]
as the \emph{honest blockrates} for hash types $i$, and 
\[
b_{i}=\frac{q_{i}^{a}}{d_{i}}
\]
as the \emph{dishonest blockrates} for hash types $i$. These quantities
are measured in blocks per second. Our security results will be written
in terms of these honest and dishonest blockrates. Since these are
proportional to the hashrate, and hashrate is proportional to the
amount of hashing resources being run, these quantities can also be
conceptualized as a sort of constrained resource. 
\begin{defn}
In a $n$ resource model, we say that the honest nodes have blockrates
$\boldsymbol{h}=(h_{1},h_{2},\ldots,h_{n})$ and the dishonest nodes
have blockrates $\text{\textbf{b=}}(b_{1},b_{2},\ldots b_{n})$. The
ordered pair of ordered n-tuples $\left(\boldsymbol{h},\boldsymbol{b}\right)$
is called the \emph{blockrate configuration. }
\end{defn}

\section{An information theoretic motivation: Can security be conditioned
on an AND of 51\% attacks?}\label{sec:An-information-theoreticMotivation}

\subsection{The motivation}

A fundamental idea of information theory is that certain events (like
erasures, bit flips, Gaussian noise, etc.) are independent. This,
combined with coding, allow for exponential decrease in probability
of error when coding is used. This was the inspiration for our approach
in this paper.

The Bitcoin algorithm is insecure only if at least $50$ percent of
hashing power is adversary. From this, we may ask, can we make Bitcoin
insecure only when $50$ percent or more of multiple hash types are
adversarial? 

For this section we treat the honest and dishonest hashrates as random
variables. This models the idea that these hashrates are unknown,
and the randomness models a user's uncertainty about the state of
the hashrates. To be clear, this randomness is \emph{not }the randomness
that arises from the hashing contests, which we analyze later.

Symbolically, let there be $n$ hash types. We define $\beta_{i}=\frac{b_{i}}{h_{i}+b_{i}}$
to be the fraction of hashing power of type $i$ that is dishonest.
We define $\theta_{i}=1-\beta_{i}$ as the fraction of mining power
of type $i$ that is honest.

Ideally, we want to construct a protocol that is secure whenever 
\begin{equation}
\bigcup_{i=1}^{n}\left\{ \beta_{i}<0.5\right\} .\label{eq:TheDesiredBigOr}
\end{equation}
If this were true, then insecurity can only occur when:
\begin{equation}
\bigcap_{i=1}^{n}\left\{ \beta_{i}\ge0.5\right\} .\label{eq:theDesiredBigAnd}
\end{equation}

We call this the ``Big-And'' security region.

Let $A_{i}$ represent each event $\left\{ \beta_{i}\ge0.5\right\} $.
If we assume each event $A_{i}$ is independent, and is each bounded
by some probability $p<1$, then if we had a protocol that satisfied
our ideal condition:
\[
P(\text{insecure})<\prod P(A_{i})\le p^{n}\le A\exp(-cn)
\]
where we use arbitrary constants $A$ and $c$ in the last expression.

Indeed, if we could design a protocol with such a security region,
we would get a much desired exponential scaling in probability of
error with the number of hash types used. Unfortunately, this is not
possible for \emph{any }permissionless protocol, and not just Merged
Bitcoin which we propose in this paper, which we prove in the next
two sections.

\subsection{Universal Conditions on Security Region}\label{sec:Conditions-on-Security}

We now consider a general result for all permissionless blockchain
protocols. We make the result very general, in order to rule out the
possibility of the ``Big-And'' security region above. Roughly speaking,
we will show that if honest and dishonest blockrates are swapped,
then a secure blockrate configuration becomes insecure. However, for
full generality, we must define ``secure'' appropriately, which
requires a subtle distinction from prior definitions of security.
In all previous definitions of security this distinction has not mattered,
generally because the proof of security of existing protocols show
that they satisfy a stronger definition anways. 

Note that in this section we do not make any strong assumptions about
network delays, or even the nature of the block arrival distributions.
We only require permisionlessness, a chain formation property, the
adversary equivalence property, and adversary withhold ability. The
theorem presented in this section indicates that if a protocol is
ever to attain the ``Big-And'' security region, at least one of
these conditions must be dropped.
\begin{defn}
A\emph{ permissionless blockchain protocol }is a protocol that uses
some resource to produce a data structure we call a \emph{blockchain,
}which has a chain of \emph{blocks, }each of which may contain a set
of transactions. Such a protocol must have a\emph{ fork choice rule},
which is a rule for choosing the \emph{canonical ledger}. The canonical
ledger is that chain of blocks which is to be considered the valid
sequence of blocks. Importantly, the fork choice rule must only depend
on the properties of the data structure produced, and not on any pre-identified
agents or information. This is the permissionless property.
\end{defn}
We also consider a general form of a blockchain protocol, with the
key property that such a protocol forms a \emph{chain}:
\begin{defn}
A blockchain protocol must have the \emph{chain }property. It must
be a linked sequence of blocks, where each link is fixed at the time
the block is produced.
\end{defn}

We also require an adversary to have at least as much strength as
the honest miners given the same mining resources:

\begin{defn}
A protocol and network model satisfies the \emph{adversary equivalence
requirement }if an adversary with resources $\boldsymbol{a}$ can
produce blockchains according to the same probability distribution
as honest nodes with resources $\boldsymbol{a}$. The adversary \emph{could
}be stronger, (for example, in the rest of this paper we consider
a case in which the adversary has no network delays), but it must
have at least the same strength given the same amount of resources.
\end{defn}
We must also define a withhold ability:

\begin{defn}
The adversary has \emph{withhold ability }if it can can produce a
block but then withhold presenting it to any other miner for an unlimited
amount of time. Once it presents its block to one honest miner, however,
other honest blocks can see it subject to network delays in the model.
\end{defn}

We now define a forever block:
\begin{defn}
A block is called a \emph{forever block} if there is some time in
which it becomes part of the canonical ledger for all agents, and
then remains in this canonical chain for all time.
\end{defn}
In this section, we need to define security precisely, and to do this
we first define a probability bounding function:
\begin{defn}
Let $f(t)\ge0$ be a function. If $f(t)$ monotically decreases and
$\lim_{t\rightarrow\infty}f(t)=0$, then we call it a \emph{probability
bounding function.}
\end{defn}
Finally, we can define security:
\begin{defn}
Starting at any time $s$, let the probability that no honest forever
block occurs in time $[s,s+t]$ be denoted by $P_{s,s+t}$. A permissionless
blockchain protocol with blockrate configuration $(\boldsymbol{h},\boldsymbol{b})$
is \emph{secure} if and only, for this configuration, there is a probability
bounding function $f$ in which $P_{s,s+t}<f(t)$ for all $s$, all
$t$, and all adversary attack strategies.\footnote{Note that this definition refers to an \emph{a priori }probability,
before any mining has been performed. This is distinguished from a
probability conditioned on the current state of the blockchain. In
the latter case, if at a particular time $s$ the adversary has already
mined a very long chain, this probability bound need not hold. Indeed,
for Merged Bitcoin, there is always \emph{some} probability that this
happens if the blockchain is run forever and the adversary gets extremely
lucky.}
\end{defn}
Note that merged bitcoin satisfies this property when in the security
region, because in any interval the probability of an honest block
occuring scales as $A\exp(-ct)$ by \cite{correctionPaper} for some constants $A$ and
$c$. 

\begin{lem}
A secure protocol has infinitely many honest blocks with probability
$1$.
\end{lem}
\begin{IEEEproof}
Since $f(t)$ approaches $0$, we can consider consecutive time intervals,
starting at $0$, of length $t_{1},t_{2},t_{3},\ldots$ in which $f(t_{i})<\frac{1}{i^{2}}.$
We consider a sequence of events, $S_{i}$, which is the event the
$i$th such interval has no forever honest blocks. We have $P(S_{i})\le f(t_{i})\le\frac{1}{i^{2}},$
which implies $\sum P(S_{i})<\sum\frac{1}{i^{2}}<\infty.$ By the
Borel-Cantelli lemma \cite{borelLemma, cantelliLemma}, the probability that $S_{i}$ occurs infinitely
often is $0$, and hence with probability $1$ there must be infinitely
many such intervals with honest blocks.
\end{IEEEproof}
\begin{thm}
\label{thm:symmetryTheorem}If $\left(\boldsymbol{a},\boldsymbol{b}\right)$
is secure then $(\boldsymbol{b},\boldsymbol{a}$) is not secure.
\end{thm}
Note that the above theorem is similar to Theorem 3.1 of \cite{minotaurFitziEtAl}, but with the distinction that herein we precisely define the random probability associated with this property. This subtle but important distinction is discussed in Section \ref{sec:DiscussionOfSecurityDefinition}.
\begin{IEEEproof}
Suppose $\left(\boldsymbol{a},\boldsymbol{b}\right)$ is secure and
$(\boldsymbol{b},\boldsymbol{a}$) is also secure. Then there are
two probability bounding functions, $f(t)$ for $\left(\boldsymbol{a},\boldsymbol{b}\right)$
and $g(t)$ for $(\boldsymbol{b},\boldsymbol{a})$. Then there is
some $T_{1}$ such that $f(t)<0.1$ for all $t>T_{1}$ and a $T_{2}$
in which $g(t)<0.1$ for all $t>T_{2}$. Choose $T=\max(T_{1},T_{2}).$

We shall now prove by contradiction.

If $\left(\boldsymbol{a},\boldsymbol{b}\right)$ is secure, then it
is secure against any type of attack. In particular it is secure against
a private mining attack in which an adversary produces a blockchain
in private, simulating exactly honest mining with $\boldsymbol{b}$
resources, and reveals its chain after time $T$, and then simply
stops mining. Call this an honest simulation attack. This is possible
due to the adversary equivalence requirement and the adversary withhold
ability. We call this case $1$. Let $N_{T}^{1}$ be the event that
at time $T$ (after the reveal), the honest chain has a forever block.
If this happens, at least one block among the honest blocks will be
in the chain forever. 

But then observe that no dishonest blocks that have occured up to
this point can be in the canonical chain forever. This is due to the
\emph{chain }property requirement. The canonical chain must be a linked
sequence of blocks. If a block in the adversary chain is also in the
canonical chain, then this block must occur before or after the honest
forever block. If it occurs before, then a block that occurred before
the honest forever block must link to it. But this cannot happen because
the adversary blocktree was mined in private. If the adversary block
occurs \emph{after} the honest forever block, then it must eventually
link to a chain containing the honest block. But this cannot happen
because the adversary was mining its own chain in private, and by
design of the attack was not linking to any of the honest blocks.

Note that the probability of no forever honest blocks occurring in
this interval is bounded by:
\[
P(\text{no forever blocks)}\le f(T)\le0.1.
\]
Thus, we have that the probability that an honest forever block occurs,
and that none of the adversary blocks are part of the canonical chain
is given by:
\[
P(N_{T}^{1})\ge1-P(\text{no forever blocks)}\ge1-0.1=0.9.
\]
Now we consider case $2$. In this case, the blockrate configuration
is $(\boldsymbol{b},\boldsymbol{a})$.

We consider the adversary doing an honest simulation attack again,
but this time the adversary has $\boldsymbol{a}$ hashing resources.
The adversary again simply simulates honest mining and reveals the
chain after time $T$. In this case, the honest miners simply produced
a tree with the same distribution as the dishonest miners did in case
1, and vice versa. Hence, by the arguments above, the probability
that the \emph{dishonest }chain has a forever block and the honest
chain has no forever block is greater than $0.9$.

But if this blockrate configuration was secure, the probability of
no forever honest blocks must be less than $0.1$. Hence, if blockrate
configuration $\left(\boldsymbol{a},\boldsymbol{b}\right)$ is secure
then $(\boldsymbol{b},\boldsymbol{a}$) is not secure. 
\end{IEEEproof}

\subsection{The AND is not possible}

The above theorem implies an immediate corallary, which is that the
AND condition on the insecurity region is not achievable by any permissionless
model. Suppose there exists a protocol that is secure whenever our
condition from $\ref{eq:TheDesiredBigOr}$ is satisfied:
\[
\bigcup_{i=1}^{n}\left\{ \beta_{i}<0.5\right\} .
\]

Suppose that a particular configuration of blockrates has $\beta_{i}<0.5$
for some block type $i$, but for a different block type $j$, $\beta_{j}>0.5$.
If our desired condition is satisfied for a hypothetical protocol,
then this configuration should be secure, because one of the block
types has less than $50$ percent adversarial power. However, if we
apply Theorem \ref{thm:symmetryTheorem} and swap the honest and dishonest
blockrates, then the new configuration of blockrates must be insecure.
But, in this case, $\beta_{j}<0.5$, which should satisfy condition
$\ref{eq:TheDesiredBigOr}$. This would imply the protocol is secure,
a contradiction. Hence, there cannot be a protocol that is secure
whenever $\bigcup_{i=1}^{n}\left\{ \beta_{i}<0.5\right\} $. Alas,
our hope of gaining an error probability that goes to zero exponentially
in number of hash types $n$ fails.

\subsection{The limited mining power, backdoor attack adversary}

Can we salvage any information theoretic scaling with multiple hash
types? Suppose instead we have a power-limited adversary. Such an
adversary could never compete on a traditional $51\%$ attack. 

In this case, one of the vulnerabilities of the Bitcoin network involves
a hardware backdoor attack (See the discussion in Chapter 10 of \cite{bitcoinStandard}).
This is when hardware manufacturers produce hardware that can be controlled
or activated remotely without knowledge of the hardware user. This
could be used to hijack hardware of otherwise honest operators. Of
course, there are many hardware manufacturers in existence. However,
with pure Bitcoin mining, there are still a limited numbers of Bitcoin
ASIC manufacturers that exist \cite{asicMinerWebsite}. Other hashing
algorithms, like Ethash, have fundamentally different hardware bottlenecks
(specifically memory bandwidth \cite{ethereumDesignJustificationWiki}),
and thus have different manufacturers.

Let us suppose that the honest score growth rate for each block type
is roughly equal.

Hence, the honest blockrate in the low $\Delta$ regime is given by
\[
\sum_{i=1}^{n}c_{i}h_{i}.
\]

Such an adversary can succeed if and only if he has greater than $50$
percent of the mining power. Since we assumed each $c_{i}h_{i}$ are
roughly equal, this requires gaining control of roughly $\frac{n}{2}$
of these hash types in order to break the protocol.

We now model our uncertainty over the possibility of hacking block
type $i$ with a probabilistic model. Let the event of hacking block
type $i$ be $K_{i}$. In our model, we assume that each of these
events are independent, and each occurs with some probability $p$
which is less than $0.5$. Then, the probability of hacking $\frac{n}{2}$
block types is bounded by the well-known tail bounds of a Bernoulli
distribution, 
\[
P(\text{insecure})\le A\exp(-cn).
\]
In this case, for this probabilistic model, Merged Bitcoin achieves
our sought after exponential scaling in number of hash types.

\subsection{Discussion of Definition of Security} \label{sec:DiscussionOfSecurityDefinition}
We distinguish two things between our definition of security and existing
definitions in the literature, in particular the definition used in \cite{minotaurFitziEtAl}. For brevity of discussion, we consider
an honest block or transaction ``getting through'' as meaning it
becomes included in the canonical chain, and stays there forever.
The first distinction involves transactions getting through compared
to honest blocks. The second involves a distinction between bounded,
and finite but unbounded attacks.

First, existing definitions, including in Minotaur, define security
in terms of the event that a \emph{transaction }get through, whereas
we consider the event that an \emph{honest block} gets through. Even
if an honest block gets through, this does not necessarily guarantee
a transaction gets through, since each block has a limit to the number
of transactions it can hold. There may be many pending transactions,
or the adversary may attack with an excess of proposed transactions.
Thus the implicit presumption that an honest block getting through
means that a transaction gets through is neither theoretically nor
practically correct.

Secondly, \cite{minotaurFitziEtAl} defines 'liveness' as a transaction getting through
after a certain wait time. In future Theorems, this definition is
amended with the phrase ``with overwhelming probability'' (see Theorem
6.1 of the Minotaur paper). Nevertheless, the authors never define
precisely what this means. Indeed, we do not believe that the results
in Theorem 3.1 (the impossibility result of \cite{minotaurFitziEtAl} comparable to
our Theorem \ref{thm:symmetryTheorem}) necessarily extend to all possible interpretations of the
term. For example, if we define ``with overwhelming probability''
to mean ``except with a probability that approaches $0$ as time
interval approaches infinity,'' we cannot extend the impossibility
result to this case. Thus, in our definition of ``security,'' we
define it to mean ``except with a probability bounded by some probability
bounding function $f(t)$, where $f(t)$ approaches $0$ as $t$ approaches
infinity.'' 

This is a subtle distinction. However, the definition precludes certain
attacks from an adversary. One such attack is the ``unbounded delay''
attack. In such an attack, the adversary can choose \emph{any} length
of time, and delay an honest block with high probability for that
length of time. Such a delay is still \emph{finite,} and thus in the
\emph{limit} as wait time goes to infinity, with probability $1$
an honest block gets through. Nevertheless, a protocol that allows
such an attack is not practically secure, for obvious reasons. More
importantly, however, is that we cannot construct an impossibility
result if the security definition allows for such unbounded attacks.
Hence, in this paper we introduce the probability bounding function
$f(t)$.

\section{Merged Bitcoin Protocol}\label{sec:Parallel-Mining-Protocol}

\subsection{Protocol Description}

We describe our protocol below, which we call \emph{Merged Bitcoin}.
Roughly, the idea is to run something similar to the Bitcoin protocol
using multiple different hashing resources instead of one.

The purpose of the protocol is to produce a \emph{blockchain}, which
is a sequence of \emph{blocks}, which are data structures containing
transactions and balances of various accounts stored on the blockchain.
Each of these blocks are linked to the previous block by a valid hash
which contains a reference to the previous block. The details of this
are described below: 
\begin{itemize}
\item We set difficulty levels for each of the $n$ hashing lotteries, $d_{i}$.
These remain fixed for all time. (We later will consider how adjustments
can be made in Section \ref{sec:Setting-difficuty-Levels}.) 
\item We assume that a genesis block is created at time $t=0$, and the
existence and hash of the genesis block is common knowledge to all
parties, honest and dishonest, including new parties. 
\item We assume all nodes proposing a transaction (though not necessarily
a block) have a public-private key pair, and they use this to produce
transactions that are broadcast to all the nodes in the protocol,
which are to be accepted into the next block if they are \emph{valid
}(\emph{i.e., }they are properly signed, and do not move any balance
to negative on the canonical chain). 
\item Honest miners must mine on the tip of the current canonical chain.
They should compute \\
$hash_{i}(prev,txs,nonce)$, for each resource
$i$, where $prev$ is a hash of the previous block that forms the
tip of the canonical chain, $txs$ are all the valid transactions
received so far that extend the tip of the canonical chain, and $nonce$
is a random number that represents a guess. When 
\[
hash_{i}(prev,txs,nonce)
\]
is less than some difficulty threshold for resource $i$, we say that
the miner has ``mined'' the next block with resource $i$. It is
to then broadcast this hash as well as its message to the public. 
\item For each block type $i$, there is an associated score $c_{i}>0$.
\item For each proposed chain, count the number $N_{i}$ of block type $i$
and compute the score:
\[
\text{score}=\sum c_{i}N_{i}
\]
\item \textbf{The fork choice rule:} Miners should mine on the chain with
the highest score. \label{point:TheForkChoiceRule}
\end{itemize}

\section{Security Region}

We consider a hypothetical honest chain that is subject to full delays
$\Delta$ from the adversary. We can directly apply Lemma 16 of \cite{correctionPaper} to show that this hypothetical chain has a time averaged
score growth rate which we denote $\lambda_{h}$. Computing and bounding
this honest score growth rate is the main subject of this paper, and
we will address this in following section.

As well, by applying properties of Poisson arrival processes, the
adversary chain has a score growth rate ($\lambda_{a}$) given by:
\[
\lambda_{a}:=\sum c_{i}b_{i}
\]

We will use Theorem 42 of \cite{correctionPaper}, to get our main results,
which we repeat here:
\begin{thm}
\label{thm:SecurityRegionTheorem}Merged Bitcoin is secure when $\lambda_{a}<\lambda_{h}$
and is not secure when $\lambda_{a}>\lambda_{h}.$
\end{thm}
The issue then becomes: what is the score growth rate of the honest
blocks? In the case of zero network delay, these growth rates are
easy to compute. As well, when all blocks have the \emph{same }score,
the growth rate is \emph{also }easy to compute.

However, in the case of bounded network delay with different scores,
this is not as easy to compute. However, we can compute upper and
lower bounds on the honest fully-delayed block growth rate, giving
us upper and lower bounds on the security regions of Merged Bitcoin.

Though closed form expressions are hard to compute, a straightforward
randomized algorithm \emph{can }estimate the growth rate. We show
the results of simulations of these growth rates and compare it to
our upper and lower bounds in Figures \ref{fig:upperAndLowerBoundsOnScoreGrowthRate}.

\section{Growth Rate Example and Simulation}\label{subsec:Bounds-on-score-grwoth-rates}

In the classic Bitcoin case, the growth rate of the fully-delayed
honest chain is \cite{sompolinskySecureHighRateTransactionBTC}
\[
\frac{h}{1+h\Delta},
\]
where $h$ is the honest blockrate for the single hash function used
in the Bitcoin protocol. However, when different blocks have different
point amounts, this growth-rate is no longer so simple to calculate.
To see why this is true, consider Figure \ref{fig:smallBlockIncreaseProcess}.
In this we compare the classic Bitcoin case with a case where some
blocks have higher score than others. We can see that for the same
arrival times, the shape of the resulting fully-delayed honest tree
is different. This is because blocks arriving within $\Delta$ of
an otherwise non-orphaned block may not necessarily make it to the
fully-delayed chain.

\begin{figure}
\includegraphics[scale=0.95]{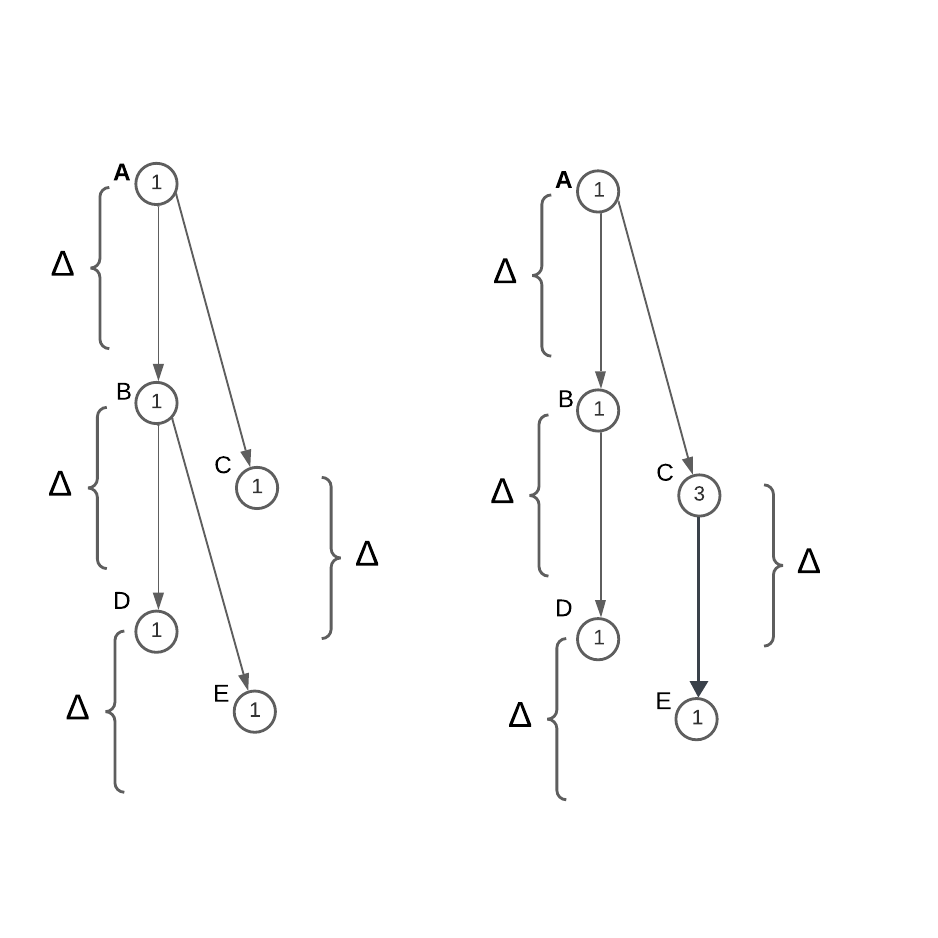}

\caption{Figure showing a fully-delayed honest chain growing when blocks have
the same point value (left) and blocks have different point values
(right). The point value of blocks are labeled with a number inside
each circle, and a letter label is on its left. The blocks are positioned
top to bottom based on the time they arrived; parentheses indicate
the time passed for one full block delay of $\Delta$. Note that when
point values are the same, block B orphans all blocks that arrive
within $\Delta$ of the block, and the chain formed by this set of
arrivals have blocks labeled (in order), ABD. However, when point
values are different, a block that arrives within $\Delta$ of a block
B may eventually be included in the final chain, and in this case
the chain formed has labels ACE. This is because the node mining block
$E$ sees a chain of score $4$ in subchain AC compared to 2 with
subchain AB.}\label{fig:growthRate}
\end{figure}

We can estimate $\lambda_{h}$ to arbitrary precision using a randomized
algorithm. Since we know the optimal delay schedule, for any $h_{i}$,
$c_{i}$, and $\Delta$, we can simulate a large number of Poisson
arrival times, compute the longest chain given the maximum delay schedule,
and then compute the score per unit time of the resulting longest
chain. It is challenging to compute a closed-form expression for $\lambda_{h}$.
However, we can use some techniques to \emph{bound $\lambda_{h}$},\emph{
}which we do in following sections.

\begin{figure}
\includegraphics[scale=0.85]{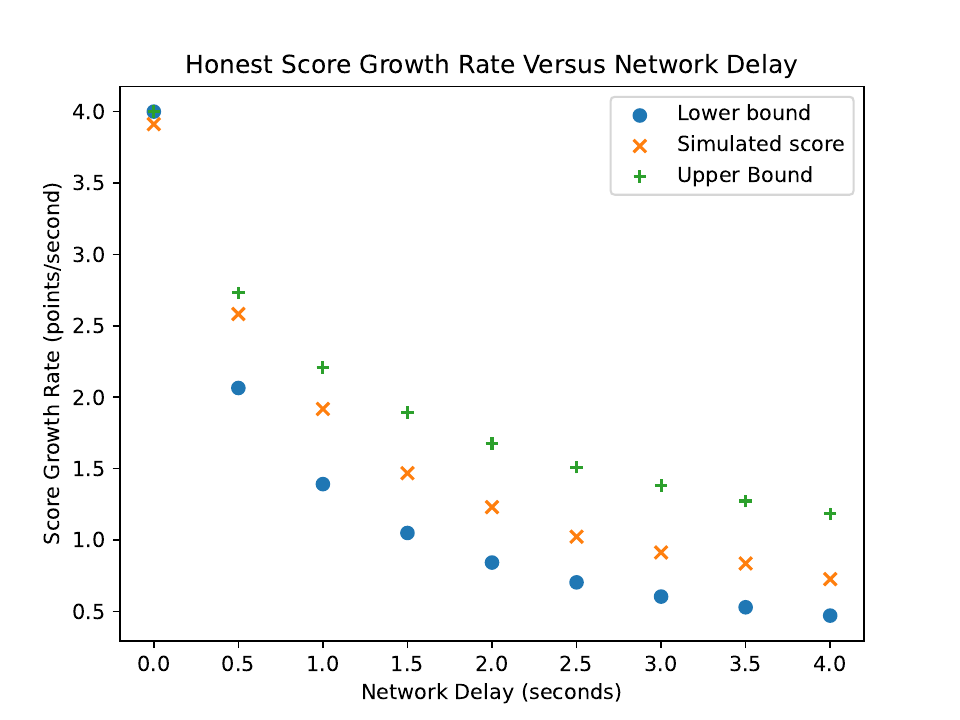}

\caption{Upper and lower bounds, as well as results from an 1000 second simulation,
for score growth rate (in points per second) when honest blockrates
are $h_{1}=2$ and $h_{2}=1$ (measured in hashes/second), and score
constants are $c_{1}=1$ and $c_{2}=2$ (measured in points per hash).
We vary the network delay $\Delta$ in increments of $0.5$ from $0$
to $5$ seconds. Note that for the $\Delta=0$ case, our upper and
lower bounds are equal and our simulated result is lower than our
lower bound due to random variation.}\label{fig:upperAndLowerBoundsOnScoreGrowthRate}
\end{figure}

We derive upper and lower bounds for the fully-delayed growth rate
in the subsections that follow, and we show a graph of these bounds
compared to a simulation in Figure \ref{fig:upperAndLowerBoundsOnScoreGrowthRate}.

\section{Simple cases of $\Delta=0$}

When $\Delta=0$ (the zero network delay case), we can see that $\lambda_{h}=\sum c_{i}h_{i}$.
This implies the following theorem: 
\begin{cor}
\label{cor:zeroDelayBound} For Merged Bitcoin with zero network delay,
if 
\[
\sum c_{i}b_{i}<\sum c_{i}h_{i},
\]
then Merged Bitcoin is secure. \label{thm:linearScoreTheorem-2} 
\end{cor}

\section{Special Case of Two Block Types}

In this section we consider the case of two block types. The block
type with the lower score we call the Small Block and the one with
the larger score we call the Big Block.

\subsection{Lower bound}\label{sec:lowerBoundSection}

\begin{figure}
\includegraphics[scale=0.75]{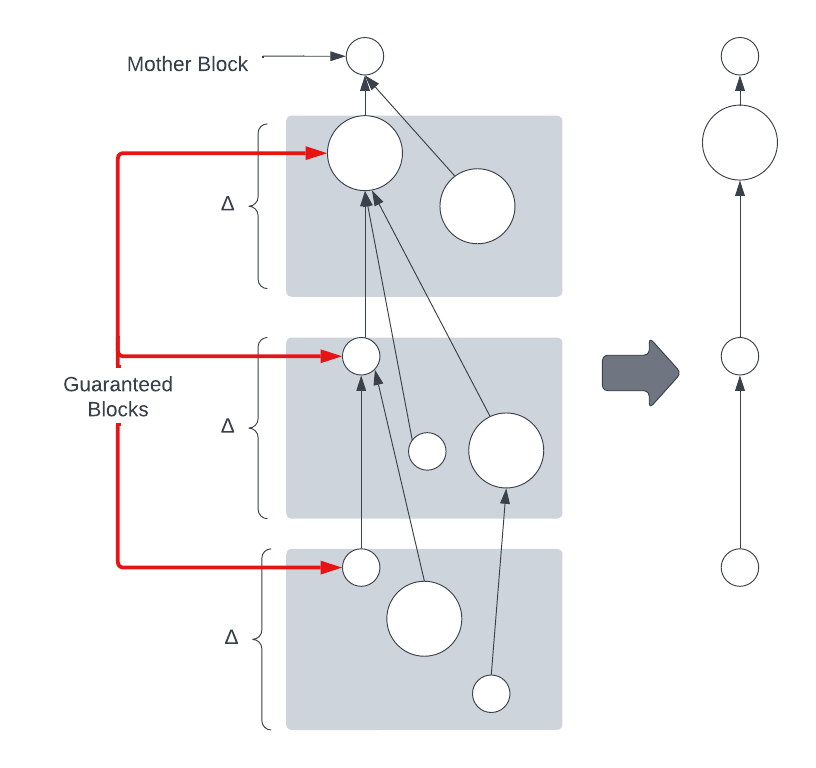}

\caption{ Figure of Delta Interval Deletion Process as applied to a particular
tree with a set of arrival times. The blocks are arranged top to bottom
based on the time of their arrival. The guaranteed blocks are labelled
and the interval of length $\Delta$ following each guaranteed block
is shaded in gray. Note that in this process all blocks that arrive
in the gray regions are deleted, resulting in the chain on the right.}\label{fig:deltaIntervalDeletionProcess}
\end{figure}

\begin{lem}
\[
\lambda_{h}\ge\frac{c_{1}h_{1}}{1+h_{1}\Delta+h_{2}\Delta}+\frac{c_{2}h_{2}}{1+h_{1}\Delta+h_{2}\Delta}.
\]
\end{lem}
\begin{IEEEproof}
For a lower bound, we use the rule that given any set of arrival times,
deleting any blocks can only keep the score of the canonical chain
the same or less (see Lemma 5, \cite{correctionPaper}). So, for this bound we consider
a deletion rule that we call the Delta Interval Deletion Process.
Let us start at the genesis block. At the next arrival, whether it
be a Big Block or Small Block, call this \emph{guaranteed block one.}
Then, delete all the blocks that arrive within time $\Delta$ of this
block. After this time, the next arrival is called \emph{guaranteed
block two}. Delete all the blocks in the time $\Delta$ following
this block. Continue this process for all time. An example of the
Delta Interval Deletion Process is illustrated in Figure \ref{fig:deltaIntervalDeletionProcess}.

The resulting chain is composed entirely of guaranteed blocks (since
all non-deleted arrivals are mined on top of a guaranteed block, since
they arrive in time at least $\Delta$ after a guaranteed block).

Let $r_{i}$ be the block arrival rate of guaranteed blocks of type
$i$ in the Delta Interval Deletion Process. 

Let $h_{i}$ be the block arrival rate for blocks of type $i$. 

We can then say that $r_{1}$ is the rate of arrival of type 1 blocks,
minus those erased in time $\Delta$ after a guaranteed block arrival.
Hence, we can say: 
\begin{equation}
r_{1}=h_{1}-r_{1}h_{1}\Delta-r_{2}h_{1}\Delta\label{eq:r1InTermsOfr1h1andDelta}
\end{equation}

In Appendix \ref{app:growthRateFormula} we make this observation rigorous for the case of
one block type. The extension to multiple block types follows the
exact same approach, which we omit for conciseness. 

Similarly, we can argue that the rate of arrival of guaranteed blocks
of type 2 is given by: 
\begin{equation}
r_{2}=h_{2}-r_{1}h_{2}\Delta-r_{2}h_{2}\Delta.\label{eq:r2RateInTermsofr1R2Delta}
\end{equation}

Rearranging the expression in \ref{eq:r1InTermsOfr1h1andDelta} give
us: 
\begin{align}
r_{1} & +r_{1}h_{1}\Delta=h_{1}-r_{2}h_{1}\Delta\nonumber \\
r_{1} & =\frac{h_{1}-r_{2}h_{1}\Delta}{1+h_{1}\Delta}\label{eq:r1Rearragned}
\end{align}
Substituting the above expression into \ref{eq:r2RateInTermsofr1R2Delta}
gives us: 
\begin{align*}
r_{2} & =h_{2}-\left(\frac{h_{1}-r_{2}h_{1}\Delta}{1+h_{1}\Delta}\right)h_{2}\Delta-r_{2}h_{2}\Delta\\
r_{2} & =r_{2}\left(\frac{h_{1}h_{2}\Delta^{2}}{1+h_{1}\Delta}-h_{2}\Delta\right)+h_{2}-\frac{h_{1}h_{2}\Delta}{1+h_{1}\Delta}\\
r_{2}\left(1-\frac{h_{1}h_{2}\Delta^{2}}{1+h_{1}\Delta}+h_{2}\Delta\right) & =h_{2}-\frac{h_{1}h_{2}\Delta}{1+h_{1}\Delta}\\
r_{2} & =\frac{h_{2}-\frac{h_{1}h_{2}\Delta}{1+h_{1}\Delta}}{1-\frac{h_{1}h_{2}\Delta^{2}}{1+h_{1}\Delta}+h_{2}\Delta}\\
r_{2} & =\frac{h_{2}(1+h_{1}\Delta)-h_{1}h_{2}\Delta}{1+h_{1}\Delta-h_{1}h_{2}\Delta^{2}+h_{2}\Delta(1+h_{1}\Delta)}\\
 & =\frac{h_{2}}{1+h_{1}\Delta+h_{2}\Delta}
\end{align*}
Substituting the above into \ref{eq:r1Rearragned} gives us: 
\begin{align*}
r_{1} & =\frac{h_{1}-r_{2}h_{1}\Delta}{1+h_{1}\Delta}\\
r_{1} & =\frac{h_{1}-\left[\frac{h_{2}}{1+h_{1}\Delta+h_{2}\Delta}\right]h_{1}\Delta}{1+\lambda_{1}\Delta}\\
r_{1} & =\frac{h_{1}(1+h_{1}\Delta+h_{2}\Delta)-h_{2}h_{1}\Delta}{\left(1+h_{1}\Delta\right)\left(1+h_{1}\Delta+h_{2}\Delta\right)}\\
r_{1} & =\frac{h_{1}(1+h_{1}\Delta)}{\left(1+h_{1}\Delta\right)\left(1+h_{1}\Delta+h_{2}\Delta\right)}\\
r_{1} & =\frac{h_{1}}{1+h_{1}\Delta+h_{2}\Delta}.
\end{align*}
Thus, the score growth rate of the deleted chain is given by: 
\[
\lambda_{del}=c_{1}r_{1}+c_{2}r_{2}=\frac{c_{1}h_{1}}{1+h_{1}\Delta+h_{2}\Delta}+\frac{c_{2}h_{2}}{1+h_{1}\Delta+h_{2}\Delta}
\]
Since the deletion method can only \emph{decrease} or \emph{maintain} the score of the chain by Lemma 5 of\cite{correctionPaper}, we have that:
\[
\lambda_{h}\ge\lambda_{del}=\frac{c_{1}h_{1}}{1+h_{1}\Delta+h_{2}\Delta}+\frac{c_{2}h_{2}}{1+h_{1}\Delta+h_{2}\Delta}.
\]
This, combined with Theorem 42 of \cite{correctionPaper}, implies a corollary: 
\end{IEEEproof}
\begin{cor}
\label{cor:lowerBound} If 
\[
\frac{c_{1}h_{1}}{1+h_{1}\Delta+h_{2}\Delta}+\frac{c_{2}}{1+h_{1}\Delta+h_{2}\Delta}>c_{1}b_{1}+c_{2}b_{2}
\]
then Merged Bitcoin is secure. 
\end{cor}

\subsection{Upper Bound}
\begin{lem}
The fully delayed honest chain growth rate, if $c_{2}>c_{1}$, is
upper bounded by: 
\[
\lambda_{h}<\frac{c_{1}h_{1}e^{-h_{2}\Delta}}{1+h_{1}\Delta+h_{2}\Delta}+\frac{c_{2}\left(h_{2}+(1-e^{-h_{2}\Delta})h_{1}\right)}{1+h_{1}\Delta+h_{2}\Delta}.
\]
\end{lem}
\begin{IEEEproof}
To produce an upper bound on the rate of growth of the honest chain,
we use the rule that \emph{increasing }the score of any block can
only \emph{increase or keep the same }the score of the canonical chain
in the view of any honest node at all times.

Without loss of generality, let us assume that $c_{2}>c_{1}$, and
we call the higher score blocks the big blocks and the lower score
the small blocks.

We consider the following process, which we call the \emph{Small Block
Increase Process}. First, we start at the genesis block. The first
block that arrives is called ``guaranteed block 1.'' The blocks
that follow in time $\Delta$ are called the orphaned blocks. Then,
after time $\Delta$, the next block to arrive is called ``guaranteed
block 2'' and those following in the next $\Delta$ are considered
orphaned blocks.

Here we are trying to find an \emph{upper bound }on the rate. Note
that if a guaranteed block is big, it indeed \emph{will }orphan any
block that arrives within $\Delta$ of it. But if a guaranteed block
is small, \emph{and a big block arrives within $\Delta$ after it,
}then it may, or may not, be orphaned in the standard protocol. However,
for the sake of our upper bound, we \emph{increase }the score of all
small blocks \emph{that have at least one big block follow it within
time $\Delta$ }to have the score of a big block. This means that
the canonical chain produced by the Small Block Increase Process will
contain only all the guaranteed blocks, with some of them having score
modified.

\begin{figure}
\includegraphics[scale=0.75]{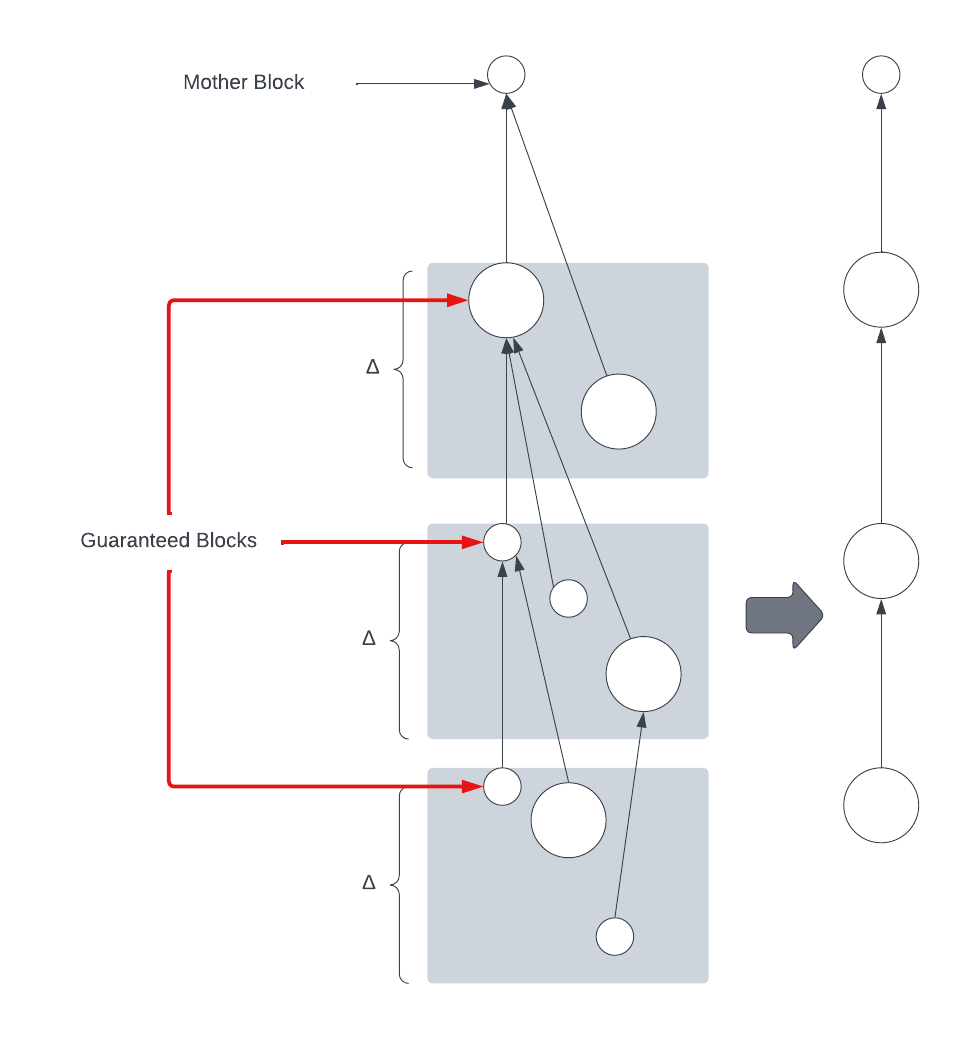}

\caption{ Figure of Small Block Increase Process. On the left is a possible
fully-delayed honest chain. Large circles are higher score blocks
and smaller circles are lower score blocks. The blocks are arranged
top to bottom based on the time they arrived. The guaranteed blocks
are labeled, and the interval of time $\Delta$ after each guaranteed
block is highlighted. To the right is what this tree is transformed
to in the Small Block Increase Process. Note that the second two guaranteed
blocks have a large block that arrives within $\Delta$ of their arrival,
so as per the rules of the process, these blocks are increased to
big blocks, forming a new chain. We prove that this new chain has
a score at least as large as the score before the increase process.}\label{fig:smallBlockIncreaseProcess}
\end{figure}

For this set of arrivals, let $\rho_{1}$ be the rate of arrival of
guaranteed small blocks, and $\rho_{2}$ be the rate of arrival of
guaranteed big blocks. (Note that whether or not a small block changes
its score to the big score, we still consider it a small block arrival).
Since these blocks grow at exactly the rates described in Section
\ref{sec:lowerBoundSection}, we have 
\[
\rho_{1}=\frac{h_{1}}{1+h_{1}\Delta+h_{2}\Delta}
\]
and 
\[
\rho_{2}=\frac{h_{2}}{1+h_{1}\Delta+h_{2}\Delta}.
\]
Now we shall consider: what fraction of the guaranteed small blocks
are going to be modified to have higher score? This is equal to the
probability that at least one big block arrives in time span $\Delta$.
Since this is a Poisson process with parameter $h_{2}$, the probability
that $0$ such blocks arrive in time-span $\Delta$ is $e^{-h_{2}\Delta}$.
Thus the probability that one or more such blocks arrive in this time
is $(1-e^{-h_{2}\Delta})$.

Let $r_{1}$ be the rate of arrival of small blocks in the Small Block
Increase Process that are \emph{not }increased. This rate is: 
\[
r_{1}=e^{-h_{2}\Delta}\rho_{1}=\frac{h_{1}e^{-h_{2}\Delta}}{1+h_{1}\Delta+h_{2}\Delta}.
\]

Let $r_{2}$ be the rate of arrival of big blocks, including those
originally big blocks, and those small blocks whose score is increased.
We have 
\[
r_{2}=\rho_{2}+\rho_{1}(1-e^{-h_{2}\Delta})=\frac{h_{2}}{1+h_{1}\Delta+h_{2}\Delta}+\frac{(1-e^{-h_{2}\Delta})h_{1}}{1+h_{1}\Delta+h_{2}\Delta}.
\]
The rate of score growth of this chain (which we denote $\lambda_{\text{incr}}$)
is given by: 
\[
\lambda_{\text{incr}}=c_{1}r_{1}+c_{2}r_{2}=\frac{c_{1}h_{1}e^{-h_{2}\Delta}}{1+h_{1}\Delta+h_{2}\Delta}+\frac{c_{2}\left(h_{2}+(1-e^{-h_{2}\Delta})h_{1}\right)}{1+h_{1}\Delta+h_{2}\Delta}.
\]
We can combine this result with Theorem \ref{thm:SecurityRegionTheorem}
to imply the following corollary: 
\end{IEEEproof}
\begin{cor}
\label{cor:upperBound} Assuming $c_{2}>c_{1},$ if 
\[
\frac{c_{1}h_{1}e^{-h_{2}\Delta}}{1+h_{1}\Delta+h_{2}\Delta}+\frac{c_{2}\left(h_{2}+(1-e^{-h_{2}\Delta})h_{1}\right)}{1+h_{1}\Delta+h_{2}\Delta}<c_{1}b_{1}+c_{2}b_{2},
\]
then Merged Bitcoin is insecure.
\end{cor}

\section{Multiple Block Types}

\subsection{Lower bound}

For the lower bound, we will extend the Delta Interval Deletion process
of Section \ref{sec:lowerBoundSection} to $n$ block types. We extend
the approach in the natural way, where guaranteed blocks are defined
as before, and we delete all blocks of all sorts in the interval $\Delta$
after a guaranteed block. Let\foreignlanguage{english}{ $r_{i}$ be
the rate of guaranteed block type $i$, and $h_{i}$ the honest block-rate
of nodes of this type. Generalizing the argument in Section }\ref{sec:lowerBoundSection}\foreignlanguage{english}{
in the natural way, we have that, for each $i\in[1,n]$:
\begin{equation}
\rho_{i}=h_{i}-h_{i}\Delta\sum_{k=1}^{n}\rho_{k}\label{eq:generalNGuaranteedBlockRateEquation}
\end{equation}
}
\begin{lem}
\label{lem:RateOfArrivalOfGuarnateedBlocksForNBlockTypes}The rate
of arrival of guaranteed blocks of type $i$, for each $i\in[1,n]$
is given by:
\[
\rho_{i}=\frac{h_{i}}{1+\Delta\sum_{k=1}^{n}h_{k}}.
\]
\end{lem}

We followed the pattern we discovered from the two block-type case
to get the theorem statement. To verify this, we plug our claimed
solution into left side of \ref{eq:generalNGuaranteedBlockRateEquation}
to get:\foreignlanguage{canadian}{
\[
\frac{h_{i}}{1+\Delta\sum_{k=1}^{n}h_{k}}.
\]
}

Plugging this solution into right side we get:
\[
h_{i}-h_{i}\Delta\sum r_{k}=h_{i}-h_{i}\Delta\sum\frac{h_{k}}{1+\Delta\sum_{j=1}^{n}h_{j}}
\]

Multiplying $h_{i}$ by $\frac{\left(1+\Delta\sum_{j=1}^{n}h_{j}\right)}{\left(1+\Delta\sum_{j=1}^{n}h_{j}\right)}$
(which is equal to $1$) gives us:
\[
\frac{h_{i}\left(1+\Delta\sum_{j=1}^{n}h_{j}\right)}{\left(1+\Delta\sum_{j=1}^{n}h_{j}\right)}-h_{i}\Delta\sum_{k=1}^{n}\frac{h_{k}}{1+\Delta\sum_{j=1}^{n}h_{j}}
\]

Simplifying:

\begin{align*}
 & =\frac{h_{i}}{\left(1+\Delta\sum_{j=1}^{n}h_{j}\right)}+\frac{h_{i}\Delta\sum_{j=1}^{n}h_{j}}{\left(1+\Delta\sum_{j=1}^{n}h_{j}\right)}-\frac{h_{i}\Delta\sum_{k=1}^{n}h_{k}}{1+\Delta\sum h_{j}}\\
 & =\frac{h_{i}}{\left(1+\Delta\sum_{j=1}^{n}h_{j}\right)}
\end{align*}

In the second to last line we recognize that the dividends in the
last two quotients differ only by the dummy variable used, and so
these two quotients sum to $0$. 

Since the left side and right side our equal, our solution is correct.

We now immediately apply this Lemma to form the following Theorem:
\begin{thm}
The honest block score growth rate for $n$ different block types
of score $c_{i}$, $i\in[1,n]$ is lower bounded by:
\[
\lambda_{h}\ge\sum\frac{c_{i}h_{i}}{1+\Delta\sum_{k=1}^{n}h_{k}}.
\]
Hence, Merged Bitcoin is secure in when:
\[
\sum\frac{c_{i}h_{i}}{1+\Delta\sum_{k=1}^{n}h_{k}}>\sum c_{i}b_{i}.
\]
\end{thm}

\subsection{Upper Bound}

Without less of generality, let's sort the block types by increasing
score, so that $c_{n}\ge c_{n-1}\ge\ldots\ge c_{1}$. We now consider
a natural generalization of the Small Block Increase Proceess. We
consider a sequence of guaranteed blocks as in the previous section.
And then for each guaranteed block that arrives and has a higher score
block that arrives within $\Delta$ of it, we increase its score to
the highest score that occurs within $\Delta$. If no higher score
block arrives within $\Delta$, we keep the same score.

Note then that the canonical chain produced by the Small Block Increase
Process will contain only all the guaranteed blocks, with some of
them having score increased. From Lemma \ref{lem:RateOfArrivalOfGuarnateedBlocksForNBlockTypes},
the growth rate of these guaranteed blocks before the increase process
is:
\[
\text{\ensuremath{\rho_{i}}=\ensuremath{\frac{h_{i}}{1+\Delta\sum_{k=1}^{n}h_{k}}}}
\]

Let us first compute the probability that a block of type $1$ is
increased in score to a block of type $2$. Denote this event $C_{1\uparrow2}$.

For this to occur, at least one block of type $2$ must occur in the
$\Delta$ interval, and no blocks of type $3$ or more arrive.

Since arrivals of each block type are independent, we can apply this
along with the well known probability mass function of a Poisson type
random variable to get:
\[
P(C_{1\uparrow2})=(1-e^{-h_{2}\Delta})e^{-h_{3}\Delta}\ldots e^{-h_{n}\Delta}
\]

In general we want to find an expression for $P(C_{x\uparrow y})$,
$x<y\le n$, which is the probability that a block of type $x$ is
followed in time $\Delta$ by a block of type $y$, but not bigger
type. We denote this probability $p_{x\uparrow y}$, and applying
independence and the standard probability mass function:\foreignlanguage{canadian}{
\[
p_{x\uparrow y}:=P(C_{x\uparrow y})=(1-e^{-h_{y}\Delta})e^{-h_{y+1}\Delta}\ldots e^{-h_{n}\Delta}.
\]
}

We shall also consider the probability that a guaranteed block of
type $i$ is \emph{not }increased. This is the probability that following
it there are no blocks that arrive within $\Delta$ of it with a bigger
score. Denote this event by $C_{x\rightarrow x}$. When $x=1$ we
have:
\[
P(C_{1\rightarrow1})=e^{-h_{2}\Delta}\ldots e^{-h_{n}\Delta}
\]
In general we denote this by the symbol $p_{x\rightarrow x}$.
\[
p_{x\rightarrow x}:=P(C_{x\rightarrow x})=e^{-\Delta h_{x+1}}e^{-\Delta h_{x+2}}\ldots e^{-\Delta h_{n}}
\]

Hence, the rate of arrivals of blocks of type $1$ in the guaranteed
block increase process is the fraction of the guaranteed blocks that
are not increased times the rate of such blocks:
\[
r_{1}=p_{1\rightarrow1}\rho_{1}
\]

For blocks of type $2$, this rate of arrival is given by the rate
of guaranteed blocks of type $2$ not increased, plus the rate of
guaranteed blocks of type $1$ that are increased to type 2:

\[
r_{2}=p_{2\rightarrow2}\rho_{2}+p_{1\rightarrow2}\rho_{1}
\]
and in general

\[
r_{k}=p_{k\rightarrow k}\rho_{k}+p_{k-1\rightarrow k}\rho_{k-1}+\ldots+p_{1\rightarrow k}\rho_{1}
\]

Now, the score growth rate of this process is given by
\[
\lambda_{\text{incr}}=c_{1}r_{1}+c_{2}r_{2}+\ldots+c_{n}r_{n}
\]

These expressions, though long, are nevertheless straightforward to
compute. We can now state the following Lemma:
\begin{lem}
Recall $h_{i}$ is the blockrate of blocks of type $i$ (before any
delays). First, let:\textbf{
\[
\rho_{i}=\frac{h_{i}}{1+\Delta\sum_{k=1}^{n}h_{k}},
\]
}

and for $x<y<n$, let:
\[
p_{x\uparrow y}:=P(C_{x\uparrow y})=(1-e^{-h_{y}\Delta})e^{-h_{y+1}\Delta}\ldots e^{-h_{n}\Delta},
\]
and 
\[
p_{n\uparrow n}=1.
\]

Also, for $x<n$, let:
\[
p_{x\rightarrow x}:=P(C_{x\rightarrow x})=e^{-\Delta h_{x+1}}e^{-\Delta h_{x+2}}\ldots e^{-\Delta h_{n}}
\]
and 
\[
p_{n\rightarrow n}=1.
\]
.

Then, in the Small Block Increase process defined above, the rate
of arrival of guaranteed blocks of type $k$ is:
\[
r_{k}=p_{k\rightarrow k}\rho_{k}+p_{k-1\rightarrow k}\rho_{k-1}+\ldots+p_{1\rightarrow k}\rho_{1}
\]
and the score growth of a fully-delayed honest chain is bounded by:
\[
\lambda_{h}\le c_{1}r_{1}+c_{2}r_{2}+\ldots+c_{n}r_{n}.
\]
\end{lem}

We can use this to directly imply a theorem for a closed form expression
for the security of Merged Bitcoin:

\begin{thm}
If 
\[
\lambda_{a}>c_{1}r_{1}+c_{2}r_{2}+\ldots+c_{n}r_{n}
\]
(with the quantities $r_{i}$ as defined above) then Merged Bitcoin
is insecure.
\end{thm}

\section{Merged Bitcoin: Increasing Cost of Optimal Attack}\label{sec:CostOfAttack}

As we can see, the feasibility of successfully attacking a merged
protocol depends on honest blockrates. Intuition says that the merged
protocols will cost more to attack, but can we make this intuition
rigorous? In this section we consider the Merged Bitcoin protocol
with zero network delay, and show that with linear costs per unit
blockrate, the cost of attacking \emph{does }increase with the merged
protocol.

Let us assume that the cost (in units of dollars/block) of hashing
power is $p_{1}$ for Hash1 and $p_{2}$ for Hash2. For comparison's
sake, let us assume that there are three options for running the distributed
protocol: using either a blockchain running the Bitcoin protocol using
Hash1 (Protocol 1), one using Hash2 (Protocol 2), or Merged Bitcoin
(Protocol 3). We note that the assumption of a linear cost per blockrate
(no matter how much hashing resource must be obtained), is a very
simplified model. Common sense and standard economic theory suggests
that the marginal cost of obtaining more of these resources should
increase as more is obtained. Nevertheless, we retain this simplified
model for illustrative purposes.
\begin{rem}
The cost of attacking (in units of dollars per second) the first two
of these protocols (pure bitcoin using either Hash1 and Hash2) is
\[
p_{1}h_{1}
\]
for Protocol 1 and 
\[
p_{2}h_{2}
\]
for Protocol 2. This flows trivially from the $50\%$ attack criteria
for the Bitcoin protocol and the assumed cost of each hashing resource.
\end{rem}
We now address the cost of attacking Merged Bitcoin assuming these
honest blockrates.
\begin{lem}
\label{lem:minimumCostOfAttacking}

Recall that $p_{1}$ and $p_{2}$ are the cost per unit hashing resource
(in dollars per block) for resources $1$ and $2$ respectively, and
$c_{1}$ and $c_{2}$ are the score constants used for the fork choice
rule in the merged protocol (measured in score per block). Assuming
the linear cost per hash resource model, the minimum cost of attacking
the protocol (in dollars per second) when network delay is zero is:
\[
\frac{p_{1}}{c_{1}}\left(c_{1}h_{1}+c_{2}h_{2}\right)
\]
if $\frac{p_{1}}{c_{1}}<\frac{p_{2}}{c_{2}}$, and 
\[
\frac{p_{2}}{c_{2}}\left(c_{1}h_{1}+c_{2}h_{2}\right)
\]
if $\frac{p_{2}}{c_{2}}<\frac{p_{1}}{c_{1}}.$ 
\end{lem}
\begin{IEEEproof}
To compute the cost of attacking the merged protocol, we combine Theorem
\ref{thm:linearScoreTheorem-2} (which state that the protocol is
insecure when $c_{1}b_{1}+c_{2}b_{2}>c_{1}h_{1}+c_{2}h_{2}$) with
our assumed cost per unit hashing resource $p_{1}$ and $p_{2}$.
This gives us the following optimization problem faced by the adversary
to minimize the cost of an attack: 
\begin{align*}
\text{minimize } & p_{1}b_{1}+p_{2}b_{2}\\
\text{subject to } & c_{1}b_{1}+c_{2}b_{2}>c_{1}h_{1}+c_{2}h_{2}
\end{align*}
Convex optimization techniques can show us that this reaches a minimum
when the adversary places all resources into the hashing resource
that, for the adversary, has lowest cost per unit score, or, in other
words, has the lowest value of $\frac{p_{i}}{c_{i}}.$ Without loss
of generality, let us suppose it is resource $1$ that minimizes this
cost. Then, the adversary minimizes its cost when: 
\begin{align*}
c_{1}b_{1} & =c_{1}h_{1}+c_{2}h_{2}\\
b_{1} & =\frac{c_{1}h_{1}+c_{2}h_{2}}{c_{1}}
\end{align*}
and this cost is: 
\[
\frac{p_{1}}{c_{1}}\left(c_{1}h_{1}+c_{2}h_{2}\right).
\]
\end{IEEEproof}
We may also ask, what values of $c_{1}$and $c_{2}$ maximize this
minimum cost? We might guess that this occurs when 
\[
\frac{p_{1}}{c_{1}}=\frac{p_{2}}{c_{2}}.
\]
Hence we can make $c_{2}=\frac{p_{2}c_{1}}{p_{1}}$ and so we might
guess that the maximum cost for attacking such a protocol is: 
\[
\frac{p_{1}}{c_{1}}\left(c_{1}h_{1}+\frac{p_{2}c_{1}}{p_{1}}h_{2}\right)=p_{1}h_{1}+p_{2}h_{2}.
\]
Thus, we see that by judiciously choosing the appropriate score function,
we can make the cost of attacking the merged protocol be the sum of
the cost of attacking each individual protocol. This does not just
maximize the cost of attacking Merged Bitcoin, it actually maximizes
the cost of attacking for \emph{any }protocol. We prove this in the
theorem below. 
\begin{thm}
\label{thm:costOptimalTheorem} If hashing resource costs are linear
and network delay is zero, by choosing $c_{1}$ and $c_{2}$ such
that $\frac{p_{1}}{c_{1}}=\frac{p_{2}}{c_{2}}$, then Merged Bitcoin
has cost of attack equal to $p_{1}h_{1}+p_{2}h_{2}$, and this is
optimal for Merged Bitcoin. Moreover, this cost of attack is the most
any multi-hash consensus protocol can attain assuming fixed honest
blockrates. 
\end{thm}
\begin{IEEEproof}
From Lemma \ref{lem:minimumCostOfAttacking}, we have that the maximum
cost of attacking Merged Bitcoin with these score constants is: 
\[
p_{1}h_{1}+p_{2}h_{2}.
\]
Suppose an adversary had slightly greater budget than this, with a
budget equal to $p_{1}h_{1}+p_{2}h_{2}+\epsilon$. Then, it can assign
its budget to obtain $h_{1}+\frac{\epsilon}{p_{1}+p_{2}}$ of resource
$1$ and $h_{2}+\frac{\epsilon}{p_{1}+p_{2}}$ of resource 2, at a
cost of 
\[
p_{1}(h_{1}+\frac{\epsilon}{p_{1}+p_{2}})+p_{2}(h_{2}+\frac{\epsilon}{p_{1}+p_{2}})=p_{1}h_{1}+p_{2}h_{2}+\epsilon.
\]
Thus, this allocation is within its budget. But the adversary \emph{also
}has more of both resource 1 and resource 2. It can then carry out
a private chain attack and, by the law of large numbers, we can be
sure that after a sufficiently long time, it will dominate the honest
chain.
\end{IEEEproof}

\section{Setting difficulty Levels}\label{sec:Setting-difficuty-Levels}

\subsection{Setting block reward and difficulty levels}

In addition to the proof-of-work based consensus mechanism, Bitcoin
also gives a reward for 

In Bitcoin, each block is given the same block reward (except, of
course, after the reward is halved according to the halving schedule
\cite{Nakamoto_bitcoin}). For Merged Bitcoin, we use this observation
to propose a rule for setting the relative difficulty of each hash.

\textbf{Proposed rule: }For Merged Bitcoin, for different types of
blocks, the block reward, the average \emph{cost} \emph{per block},
and the score \emph{per block }(defined as $c_{1},c_{2}$ above) should
all be equal.

Recall that $d_{i}$ is the difficulty setting for $hash_{i}$ (measured
in hashes per block). We let $\kappa_{i}$ be the cost per hash for
$hash_{i}$ (measured in dollars per hash). The \emph{cost per block
}quantities we define below: 
\[
C_{1}=d_{1}\kappa_{1},C_{2}=d_{2}\kappa_{2}.
\]
These are measured in units of (dollars/block). The criteria we can
use applying our proposed rule is thus: 
\begin{align}
d_{1}\kappa_{1} & =d_{2}\kappa_{2},d_{2}=d_{1}\frac{\kappa_{1}}{\kappa_{2}}.\label{eq:ruleForDifficultySetting}
\end{align}

\subsection{Relative Difficulty Adjustment}

In the previous sections we have analyzed the protocol assuming constant
difficulty thresholds for each hash type. In the above section we
showed a principled way to select the relative difficulties of the
hashes. We can also adjust the absolute difficulties of the hashes
just as Bitcoin is adjusted: the difficulty level gets adjusted in
the long run so that the rate of arrival of new blocks stays relatively
constant.

However, when there are two different hashes there is one problem
facing the blockchain in the long term that does not face Bitcoin:
suppose that through technological advance one type of hash becomes
substantially cheaper than the other one? This has substantial impacts
on the composition of the merged blockchain. Specifically, if the
relative difficulty threshold in the protocol of each hash stays the
same, but one hash becomes much cheaper to produce, and the block
rewards stay the same for each hash type, then it may emerge that
all or close to all of the blocks are produced with the cheaper hash.

To remedy this, we propose a simple rule for difficulty adjustment
to prevent this issue. We call it \emph{minimum fraction difficulty
adjustment.}

\textbf{Proposal}: For adjusting relative difficulties in the protocol,
if the fraction of one hash type goes below a threshold in terms of
fraction of blocks produced, it's relative difficulty should be incrementally
adjusted down (while keeping the block reward the same).

We make the following assumptions about the behaviour of agents in
a merged blockchain: 
\begin{itemize}
\item There is some initial honest mining power for each hash function,
and some fraction of this honest mining power is used in the merged
model. 
\item The costs of obtaining (and running) more mining power vary per user
and vary over time, but remain relatively close to the average market
cost. 
\item In the medium to long term, miners of a particular type will join
if mining is profitable and they will exit in the long run if mining
is unprofitable. 
\item Mining is profitable if block reward exceeds the cost of mining, and
if mining is profitable, then there will always be a minimum fraction
of honest nodes participating in the protocol for each hash type. 
\end{itemize}
\begin{rem}
In the long run, Merged Bitcoin with minimum fraction difficulty adjustment
will be composed of at least a minimum fraction of nodes of each type.
To see this, suppose that the fraction of one type of block goes below
a threshold. Then, by the rules of minimum fraction difficulty adjustment,
the relative difficulty of this hash type will be adjusted down. This
will keep happening until the cost of producing a block of this type
is below the block reward. By the assumption above, miners will then
join using this hash type because it is now profitable.

Difficulty adjustment may be a vulnerability in the protocol. In particular,
targeted mining of a particular type could allow for a difficulty
adjustment that gives an adversary an advantage not considered in
the security arguments referenced in this paper. Proving security
region bounds for Merged Bitcoin with difficulty adjustment remains
an area of future work.
\end{rem}

\section{Comparison of Minotaur with Merged Bitcoin}\label{subsec:Comparison-of-Minotaur}

The closest comparable protocol to Merged Bitcoin is Minotaur \cite{minotaurFitziEtAl}.
This is a merged proof-of-work and proof-of-stake protocol. This protocol
uses a proof of stake backbone, either Ourobouros \cite{ourobourosAggelosKiayias},
Ourobouros Praos \cite{OurobourosPraos}, or Ourobouros Genesis \cite{ourobourosGenesis}.

An easy adaptation of this protocol can convert it into pure merged
proof-of-work by granting stake to proof-of-work miners in proportion
to the block counts in prior epochs. The security region of such a
merged protocol could achieve similar security regions as Merged Bitcoin,
with significant caveats. First, there is a always a chance of a chain
outage, in which the protocol as specified can no longer run. In addition,
Minotaur has two fundamental problems: the \emph{takeover attack}
and the \emph{prospective attack. } We describe the outage issue and
the two weaknesses in the following subsections.

\subsection{Minotaur Chain Outage}

If by some chance \emph{no} blocks are produced in the recency interval
by either honest or dishonest nodes, then the Minotaur chain will
simply end. This does not occur with Merged Bitcoin, and is a fundamental
problem with the Minotaur protocol. This probability can be made very
low, but with probability one this eventually occurs. This is in contrast
with the Bitcoin protocol, which can always continue so long as some
nodes eventually continue mining.

\subsection{Minotaur takeover attack}

There is a substantial asterisk on security of Minotaur as claimed
in \cite{minotaurFitziEtAl}. Fundamentally, the authors are using
a security definition that is weaker than the security definition
we are claiming for Merged Bitcoin. For a protocol to be useful, it's
long term fraction of honest blocks (the chain quality), must be above
$0$. (See the definition of Chain Quality, Definition 3, in \cite{BackboneProtocolGaray}).
In this section we explain how the chain quality of Minotaur is actually
$0$ with probability $1$ (so long as an adversary has \emph{any
}mining power).

A parameter of the Minotaur protocol is the time from when a block
is produced to the latest time it can be referenced. Let's call this
the \emph{recency interval.}. Minotaur works using a proof-of-stake
backbone where stake is granted to proof-of-work miners in proportion
to the number of proof-of-work blocks which are mined and \emph{also
}referenced by a proof-of-stake block that is produced within the
recency interval.\emph{ }

This gives rise to a fundamental problem. If, by coincidence, \emph{all
}the proof-of-stake blocks mined are dishonest nodes for an entire
recency interval, then the adversary can reference \emph{only }their
own proof-of-work blocks. From then onward the adversary can fully
control the chain, even if it has a substantially smaller amount of
resources.

The probability that this occurs in any particular\emph{ }interval
can be made very small, but blockchains run forever. Any adversary
with any non-zero resource amounts will \emph{eventually }succeed,
because the recency interval is finite. Thus, the long term fraction
of honest blocks for Minotaur is $0$, with probability $1$. This
is in contrast to Merged Bitcoin, which we prove in this paper has
fraction of honest blocks above $0$ for all time, with probability
$1$, for all points in the security region.

This is \emph{not} just a theoretical or academic issue that would
never arise in practice. Any short term outage of a sufficient fraction
of honest nodes can make the probability of a takeover attack likely.
In contrast, short term outages of honest nodes in Merged Bitcoin
can result in reversal of transactions in the short term (a security
issue), but such issues are temporary and last only as long as the
outage.

Moreover, if the honest and dishonest blockrates are \emph{close }to
the security area boundary as defined in Minotaur, there is also a
high chance that within a reasonable time the adversary can successfully
carry out a takeover attack. In contrast to this, for \emph{every
}point in the security region of Merged Bitcoin, the chance that the
adversary can ever take over the chain is precisely $0$, even if
this point is very close to the security region boundary.

\subsection{Prospective Attack}

Adaptive attacks involve an adversary knowing in advance which nodes
(other than the adversary itself) are going to form future blocks.
This allows the adversary to offer these nodes targeted bribes in
order to take over the protocol. Unlike an adaptive attack, which
Ourobouros Praos \cite{OurobourosPraos} (and hence Minotaur) resolves,
a \emph{prospective attack} involves the adversary knowing in advance
which blocks the adversary \emph{itself} will produce.

This could easily be resolved with a reliable proof-of-time combined
with the proof-of-stake backbone. This is usually done with a verifiable
delay function. However, proofs-of-time may be subject to an adversary
producing a specialized circuit that solves the verifiable delay functions
significantly faster than the fastest honest time, making this approach
potentially unreliable. On the other hand, with Bitcoin, as well as
with Merged Bitcoin, the adversary does not know when they will produce
a block \emph{until it is actually mined. }

To see why this matters, let us consider Minotaur and Bitcoin, with
parameters set so that arrival times of honest and dishonest blocks
are comparable. Then consider a store that
accepts blockchain-based tokens as payment in exchange for a tangible
good like collector coins. For practical purposes, the store owner,
who we shall call Alice, will accept the payment after four block
confirmations after the payment has been sent.

How could this be attacked in the Minotaur case? Well, an adversary,
who we shall call Bob, who gets lucky once in a while, sometimes will
produce a side chain that has a blockrate substantially better than
average. Usually this doesn't happen, but sometimes it must. The key
point is that by precomputing the hashes, Bob is able to know \emph{in
advance }whenever he has a prospective side chain that is likely to
outcompete with the honest chain.

Bob can simply go to the physical coin store when he knows his upcoming
blockrate is untypically high (and likely to orphan the honest chain).
He can generate a private chain where he moves all his money into
a new private wallet, and then in public can send his payment to the
coin store. After four public confirmations of this transaction, Alice
will give Bob the physical coins. Then, Bob can leave the store, and
produce the privately-mined, abnormally-high-score private chain,
orphaning the chain containing the payment for the physical coins,
hence stealing the coins.

Consider trying to do this with Bitcoin, or Merged Bitcoin. Before
a transaction takes place, Bob has no idea whether in the future his
blockrate will be abnormally fast, and thus Bob \emph{does not know
in advance if his sidechain will dominate.} So, to carry out the same
style of attack with Bitcoin or Merged Bitcoin, Bob would have to
go to the store \emph{every time }he tries to mine a side chain. This
costs him in multiple ways that are not present with the comparable
attack on Minotaur: 
\begin{enumerate}
\item He must pay for the mining of a side chain (which costs considerable
mining expenses), and risk not winning any block rewards for that
effort. 
\item He must go to the store physically each time and buy the physical
coins; this costs time and money \emph{each time.} 
\item He must pay transaction fees each time he goes to the store to attempt
a double spend attack. 
\end{enumerate}
Because of this, Bob may decide that attempted double spend attacks
are never worth it with Merged Bitcoin, whereas with Minotaur it costs
nothing to \emph{try }a side chain attack, and then only attempt the
double spend when he is likely to succeed. Moreover, Alice may choose
to accept a token built on Bitcoin (or Merged Bitcoin), but refuse
to accept one based on Minotaur, even though the security regions
may be similar for both.

\section{Other Advantages of Merged Mining}\label{sec:AdvantagesOfMergedMining}

\subsection{Coordination of dishonest players of different hash types more difficult
than honest coordination}

Could a similar advantage to Merged Bitcoin be gained by simply increasing
the honest hashing power? Hypothetically, the cost of attack could
be increased without limit. But one advantage of Merged Bitcoin is
that it allows honest players to secure a protocol without outside
coordination, but dishonest players must coordinate.

Let $h_{1}^{i}$ and $h_{2}^{i}$ be the initial honest blockrates
of all the hashing resources used for hash types $1$ and $2$ respectively.
(For example, they could be SHA-256 miners and Ethhash miners). If
these nodes participate in Merged Bitcoin, when there is zero network
delay: 
\[
\lambda_{h}=c_{1}h_{1}^{i}+c_{2}h_{2}^{i}
\]
Let's suppose that the initial dishonest blockrates for these hash
types are $b_{1}^{i}$ and $b_{2}^{i}$. Can we simply add these up
as we did with the honest resource amounts to compute the initial
dishonest score growth rate?

Not exactly. Dishonest players must coordinate with one another in
order to attack the chain. These dishonest resources have different
histories and different economic motivations. First, it may not be
possible to coordinate these players. Attempting to coordinate among
the dishonest players could risk revealing the attempted attack, damaging
the ability to successfully carry out a double spend attack against
an unwitting participant. This puts the merged protocol at a strategic
advantage.

\subsection{Robustness to asymmetric adversary advantages in hashing due to algorithmic
or quantum advances}

The Bitcoin protocol also has another potential vulnerability. There
may be algorithmic or circuit design advantages that an adversary
could gain for certain types of hash functions.

One such vulnerability is a quantum attack. Note that attacking the
proof-of-work mechanism of Bitcoin is distinct from attacking the
digital signature. The latter will be vulnerable to a Shor's algorithm
attack \cite{shorsAlgorithm}, but the former is not expected to have
any exponential speedup with quantum algorithms \cite{navigatingQuantumThreat}. Nevertheless,
the well-known Grover's algorithm \cite{groversAlgorithm} could give
an adversary a square-root speedup in the time to find a hash. This
is sufficient for a well-equipped adversary to completely dominate
the chain.

Merged Bitcoin could provide a partial remedy for this. Grover's algorithm
requires a quantum unitary operator that implements the particular
hash function. Each such quantum hash function may require its own
specialized quantum circuit. Thus, if quantum computing emerges as
something practical, there may be a period where an adversary could
gain an advantage on one type of hashing contest. But getting an advantage
on multiple types of hashing resources at a time may require simultaneously
having multiple specialized quantum computers, which could exceed
their capabilities. Thus, Merged Bitcoin can provide a remedy to temporary
asymmetric advantages the adversary may have due to a quantum computer.

Note that in the long run, if honest and dishonest players have access
to quantum Grover's algorithm hashes, then proof of work hashing does
not have a fundamental vulnerability. This is because this algorithm
only gives a square root speed up in number of operations required
(and so with a sufficient difficulty threshold blockrates could be
kept suitably slow).

\subsection{Broader trust across different types of users}

A separate advantage of merged mining is allowing for agreement between
two parties with different trust levels for different hash types.
This is best illustrated in an example.

Consider two parties, the Bitcoin maximalist and the Ethereum maximalist.
Let's also assume that the total mining power of each type is $1$,
and that score constants are $1$. The Bitcoin maximalist considers
the Bitcoin miners very trustworthy, and believes that at most up
to $0.1$ of these miners are dishonest, however he thinks that up
to 0.6 of the Ethereum hash power may be corrupted. We denote these
estimates, respectively, as follows: 
\[
\tilde{b_{1}^{B}}=0.1,\tilde{b_{2}^{B}}=0.6
\]
The Ethereum maximalist is inverted, and trusts the Ethash miners
more. She estimates the corruptible fractions of the Bitcoin and Ethereum
mining power to be, respectively: 
\[
\tilde{b_{1}^{E}}=0.6,\tilde{b_{2}^{E}}=0.1
\]
For this example, the Bitcoin maximalist cannot trust a blockchain
run on Ethereum hashes, and the Ethereum maximalist cannot trust a
chain run on Bitcoin hashes. However, in the view of both of these
parties, the merged chain will have only $0.6+0.1=0.7$ dishonest
mining power, and the honest will have $1.3$. Thus, Merged Bitcoin
would be trusted by both parties.

\section{Conclusions and Future Work}\label{sec:Conclusion}

In this paper we introduced Merged Bitcoin. We argue that by allowing
for multiple different hash types, Merged Bitcoin allows for greater
decentralization and higher cost to attack.

The main technical work of the paper was an analysis of the security
regions for the $\Delta$ bounded network delay model. For this model,
we found closed-form expressions for lower and upper bounds on the
security region, and for the case of equal score constants and zero
network delay, we derive exact closed-form expression for the security
region.

We argued that, unlike with a competing merged protocol Minotaur,
Merged Bitcoin does not face the problems of \emph{prospective attacks}
or \emph{takeover attacks.} We also showed that Merged Bitcoin can
maximize the cost of attacking a protocol in the linear-cost-per hash
model. We argued it was robust against asymmetric advantages an adversary
may gain due to quantum attacks, and argued that Merged Bitcoin could
allow for broader trust among different types of users. 

We showed that no permissionless blockchain protocol that allows adversary
withholding can be constructed which precludes the security region
from being the ``Big-And'' of a collection of 51\% attacks. However,
a comparable exponentially small error probability can be maintained
against a power-limited adversary trying to conduct multiple hardware
backdoor attacks. 

We are still interested in merged protocols that have non-linear security
regions (like those found in Minotaur). We also are interested in
protocols that can reach the ``Big-And'' security region of Section
\ref{sec:An-information-theoreticMotivation} if some of the assumptions
about adversary power are dropped. (For example, if the adversary
cannot withhold its blocks, a ``Big-And'' security region may be
achievable). Analyzing the $k$-confirmation rule for Merged Bitcoin
is also an area of future work. 

\bibliographystyle{IEEEtran}
\bibliography{bibtextDoc}

\appendices
\section{Growth Rate Formula}\label{app:growthRateFormula}
We will prove here a simple case, in which we show that in the single
block case, (the standard Bitcoin case)
\[
\lambda=h-\lambda\Delta h
\]
where $\lambda$ is the rate of arrival of guaranteed blocks, and
$h$ is the blockrate of these blocks before delay.

Let $t$ be time, and let $N(t)$ be the total number of honest blocks
in the fully delayed chain by time $t$. Let $\lambda$ be defined
as the random variable:
\begin{equation}
\lambda=\lim_{t\rightarrow\infty}\frac{N(t)}{t}\label{eq:definitionOfNt}
\end{equation}
(which we know from REF is some constant with probability $1$).

We let $H(t)$ be the number of blocks (in a full chain without delays)
that have arrived by time $t$, and $D(t)$ the number of orphaned
blocks by time $t$. We have that:

\[
N(t)=H(t)-D(t)
\]

Hence, 
\begin{equation}
\lim_{t\rightarrow\infty}\frac{N(t)}{t}=\lim_{t\rightarrow\infty}\frac{H(t)}{t}-\lim_{t\rightarrow\infty}\frac{D(t)}{t}\label{eq:TheLimitEquation}
\end{equation}

The left side is $\lambda$ by definition, and 
\begin{equation}
\lim_{t\rightarrow\infty}\frac{H(t)}{t}=h\label{eq:PoissonDefinition}
\end{equation}
with probability $1$, by the strong law of large numbers.

Let $T_{orphaned}(t)$ be the total time in all $\Delta$ time interval
up to time $t$. This means that, $T_{orphaned}(t)=(N(t))\Delta-\Delta c(t)$
where $c(t)$ is a random variable less than $1$, representing the
fraction of time after a guaranteed block that is cut off by our time
$t$ . Note that $D(t)$ has the same distribution as a Poisson process
that has run for time $T_{orphaned}(t)$. Call this other Poisson
process $H'(T_{orphaned}(t))$. Hence:

\begin{align*}
\lim_{t\rightarrow\infty}\frac{D(t)}{t} & =\lim_{t\rightarrow\infty}\frac{H(T_{orphaned}(t))}{t}=\lim_{t\rightarrow\infty}\frac{H(T_{orphaned}(t))}{T_{orphaned}(t)}\frac{T_{orphaned}(t)}{t}\\
= & \lim_{t\rightarrow\infty}\frac{H(T_{orphaned}(t))}{T_{orphaned}(t)}\lim_{t\rightarrow\infty}\frac{T_{orphaned}(t)}{t}\\
= & h\lim_{t\rightarrow\infty}\frac{\left((N(t))\Delta-\Delta c(t)\right)}{t}\\
= & h\lambda\Delta
\end{align*}

In the second line we restrict ourselves to considering the limits
when these two limits exist and $T_{orphaned}(t)$ approaches infinity
(which by law of large numbers each occur with probability $1$, so
their and also occus with probability $1$). In the third line we
use the fact that $\frac{H(T_{orphaned}(t))}{T_{orphaned}(t)}$ is
just the average number of arrivals of a Poisson process in time $T_{orphaned}$,
which is $h$. In the last line we use the definition of $\lambda$
from \ref{eq:definitionOfNt} and evaluate the limit $\lim_{t\rightarrow\infty}\frac{\Delta c(t)}{t}=0$

Combining this with \ref{eq:TheLimitEquation} and the limit evaluated
in \ref{eq:PoissonDefinition}means that with probability one: 

\[
\lambda=h-\lambda\Delta h.
\]

There is a natural and obvious extension of this approach to the Merged
Bitcoin case with multiple block types which we omit here for brevity.
\end{document}